# Multiferroic Properties of Electrospun $CoFe_2O_4$ – $(Ba_{0.95}Ca_{0.05})(Ti_{0.89}Sn_{0.11})O_3$ Nanocomposites for Magnetoelectric and Magnetic Field Sensing Applications

Youness Hadouch [1,2,3] *, Nayad Abdallah[4], Daoud Mezzane[2,3], M'barek Amjoud[3], Voicu Dolocan[5], Khalid Hoummada[5], Nikola Novak[1], Anna Razumnaya[1], Brigita Rozic[1], Val Fisinger[6], Hana Ursic[6], Valentin Laguta[7], Zdravko Kutnjak[1], Mimoun El Marssi[2].

**1** Condensed Matter Physics Department F5, Jozef Stefan Institute, Jamova Cesta 39, 1000 Ljubljana, Slovenia.

**2** Laboratory of Physics of Condensed Matter (LPMC), University of Picardie Jules Verne, Scientific Pole, 33 rue Saint-Leu, 80039 Amiens Cedex 1, France.

**3** Laboratory of Innovative Materials, Energy and Sustainable Development (IMED), Cadi-Ayyad University, Faculty of Sciences and Technology, BP 549, Marrakech, Morocco.

**4** The High Throughput Multidisciplinary Research Laboratory (HTMR) Laboratory, University Mohammed VI Polytechnic (UM6P), 43150 Ben Guerir, Morocco.

**5** Aix-Marseille University - CNRS, IM2NP Faculty of Sciences of Saint-Jérôme case 142, 13397 Marseille, France.

**6** Electronic Ceramics Department K5, Jozef Stefan Institute, Jamova Cesta 39, 1000 Ljubljana, Slovenia.

**7** Institute of Physics AS CR, Cukrovarnicka 10, Prague, 16200, Czech Republic.

*__Corresponding author:__
E-mail: hadouch.younes@gmail.com; youness.hadouch@ijs.si
ORCID: https://orcid.org/0000-0002-8087-9494
Tel: +212-6 49 97 06 74

## Abstract

Multiferroic $CoFe_2O_4$–$Ba_{0.95}Ca_{0.05}Ti_{0.89}Sn_{0.11}O_3$ composite nanofibers (CFO–BCTSn NFs) were synthesized using a sol-gel electrospinning method. Scanning electron microscopy revealed the morphology of the composites, with fiber diameters ranging from 120 to 150 nm. Transmission electron microscopy confirmed the structure of the nanofibers, while X-ray diffraction, Raman spectroscopy, and high-resolution transmission electron microscopy verified the formation of the spinel structure of CFO and the perovskite structure of BCTSn, with no additional phases detected. The magnetic properties of the CFO–BCTSn NFs were demonstrated by magnetic hysteresis loops (M-H), and piezoresponse force microscopy confirmed their piezoelectricity. Magnetoelectric coupling was evidenced by comparing the M-H hysteresis loops of electrically poled and unpoled CFO–BCTSn NFs samples. These composite nanofibers have the potential to be utilized in innovative, lead-free magnetoelectric and magnetic field sensing technologies at the nanoscale.

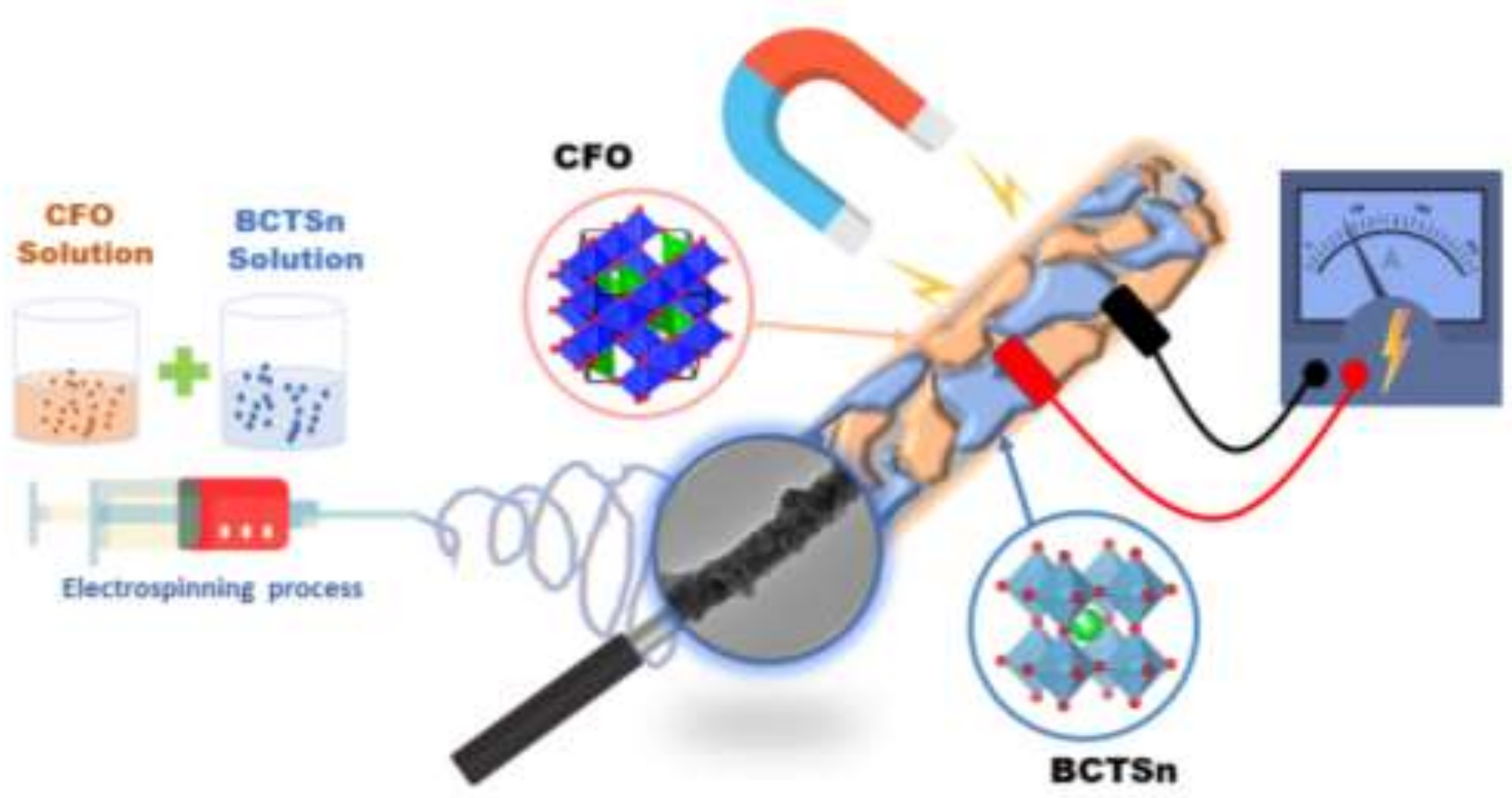

**Keywords:** *multiferroic, nanocomposite, nanofibers, electrospinning, spinel, perovskite, interface, magnetoelectric coupling.*

## Compliance with Ethical Standards:

Not applicable

## Competing Interests:

Not applicable

## Research Data Policy:

Not applicable

## Data Availability Statements:

Not applicable

## Author Contribution:

**Youness Hadouch**: Investigation, methodology, data curation, writing - Original Draft, validation,

**Nayad Abdallah**: formal analysis, data curation, validation, original draft,

**Daoud Mezzane**: supervision, conceptualization, writing – review & editing validation,

**M'barek Amjoud**: supervision, conceptualization, writing – review & editing, validation,

**Voicu Dolocan**: Formal analysis, review & editing, validation,

**Khalid Hoummada**: Visualization, review, validation,

**Nikola Novak**: Formal analysis, visualization, review, validation,

**Anna Razumnaya**: Formal analysis, visualization, review, validation,

**Brigita Rozic:** Formal analysis, visualization, review, validation,

**Val Fisinger:** Formal analysis, review, validation,

**Hana Ursic:** Formal analysis, writing–review & editing, validation,

**Valentin Laguta:** Formal analysis, visualization, review, validation,

**Zdravko Kutnjak**: Visualization, writing – review & editing, validation, supervision,

**Mimoun El Marssi**: Visualization, Writing – review & editing, validation, supervision.

## INTRODUCTION:

One-dimensional (1D) nanomaterials, such as nanotubes, nanowires, nanofibers, and nanobelts, have attracted increasing attention over the past decade due to their unique properties compared to their bulk or particle counterparts [1], [2], [3], [4]. The distinctive features, including a large surface area, high aspect ratio, and efficient electron transport pathways, make them highly promising for applications in energy storage, catalysis, and magnetoelectric fields [5], [6].

For instance, in magnetoelectric (ME) materials, 1D multiferroic materials have recently gained significant interest because they offer a much tighter coupling between ferroelectric and ferromagnetic phases, providing additional degrees of freedom in controlling size, interface, and epitaxial strain to enhance the ME coupling [7], [8]. Materials with strong ME coupling offer a variety of novel application prospects, such as dual electric and magnetic field tunable signal processing devices, very sensitive magnetic sensors, and advances in energy harvesting and information storage technology [9], [10], [11].

Since the magnetoelectric property relies on the transfer of strain between the two ferroelectric and ferromagnetic phases, creating 1D interfaces at the nanoscale emerges as the most effective approach to enhance contact between the two phases in multiferroic materials [12], [13], [14]. Various methods, such as template-assisted approaches, hydrothermal techniques, and electrospinning methods, have been utilized for the fabrication of 1D multiferroic nanocomposites [15], [16], [17].

Electrospinning is considered as a versatile method for the synthesis of 1D nanomaterials, offering practical advantages such as the ability to control the nanostructures' size, achieve high aspect ratio, with low-cost efficiency. Additionally, it is possible to easily control the properties of nanofibers using electrospinning parameters like the electric field, solution viscosity, humidity, and annealing temperature [17], [18]. Several new ideas and concepts have been explored to develop multiferroic material composites at the nanoscale using electrospinning: (i) core-shell nanofibers prepared using a coaxial needle [18], [19], [20], (ii) composite nanofibers created by electrospinning a piezoelectric/magnetostrictive solution containing fibers/particles of magnetostrictive/piezoelectric materials [21], and (iii) nanofibers produced by electrospinning a mixed solution of piezoelectric and magnetostrictive precursor solutions [22], [23], [24], [25]. In this latter method, the incorporation of two-phase ionic components in the precursor nanofibers results in a uniform blend of ferroelectric and ferromagnetic nanocrystals. This integration enhances the contact between the two phases, thereby improving magnetoelectric properties. Additionally, higher piezoelectric and magnetostriction coefficients can enhance the magnetoelectric coupling coefficient. As numerous studies have reported, $CoFe_2O_4$ emerges as the optimal candidate for the magnetic phase of multiferroic composites, demonstrating a substantial magnetostriction coefficient of approximately 200 ppm [26], [27], [28]. Conversely, lead-free perovskite has garnered significant attention due to its remarkable piezoelectric coefficient at room temperature. For instance, $0.5Ba(Zr_{0.2}Ti_{0.8})O_3$–$0.5(Ba_{0.7}Ca_{0.3})O_3$ (BCZT) exhibits a high piezoelectric coefficient ($d_{33}$ = 620 pC $N^{-1}$) due to the presence of a morphotropic phase boundary (MPB) [29]. Furthermore, $(Ba_{0.95}Ca_{0.05})(Ti_{0.91}Sn_{0.09})O_3$, featuring a pseudo-cubic Pc-O phase boundary, showcases an impressive $d_{33}$ of 670 pC $N^{-1}$ at room temperature [30].

$BaTiO_3$ (and its derivatives)/CFO multiferroic composites have been extensively studied in the literature using various connectivity types, such as (0-3) [31], [32], [33], [34], (2-2)[35], [36], [37], [38], [39], and core-shell [21], [40], [41], [42], [43]. Recently, our research reported the multiferroic properties of $Ba_{0.95}Ca_{0.05}Ti_{0.89}Sn_{0.11}O_3$–$(x)CoFe_2O_4$ (BCTSn–CFO) particulate composites (0-3), but unfortunately, the magnetoelectric properties fell short of our expectations [44]. Subsequently, we synthesized and characterized multiferroic $CoFe_2O_4$–$Ba_{0.95}Ca_{0.05}Ti_{0.89}Sn_{0.11}O_3$ core-shell nanofibers (CFO@BCTSn NFs) using a sol-gel co-axial electrospinning technique. The obtained results revealed a remarkable ME coefficient of approximately 346 mV $cm^{-1}$ $Oe^{-1}$ at a field strength of 10 kOe [20]. Based on these findings, it is concluded that manipulating the shapes and sizes of the piezoelectric and magnetostrictive phase grains has the potential to greatly enhance magnetoelectric coupling.

Our study in this paper falls within this framework, serving two main objectives. First, it aims to fabricate multiferroic $CoFe_2O_4$–$Ba_{0.95}Ca_{0.05}Ti_{0.89}Sn_{0.11}O_3$ composite nanofibers (CFO–BCTSn composite NFs) using the electrospinning method. Second, it intends to investigate the structural, morphological, piezoelectric, magnetic, and magnetoelectric properties of the synthesized CFO–BCTSn nanofibers.

## 2. EXPERIMENTAL SECTION

## 2.1. Material Synthesis

Iron nitrate nonahydrate [$Fe(NO_3)_3.9H_2O$, Oxford] (≥99.0%), cobalt nitrate hexahydrate [$Co(NO_3)_2.6H_2O$, Alfa Aesar] (≥98.0%), barium acetate [$Ba(CH_3COO)_2$ Merck] (≥99.0%), calcium acetate [$Ca(CH_3COO)_2$, Loba Chemie] (≥97.0%), tin chloride dihydrate [$SnCl_2.2H_2O_2)_4$, Fluka] (≥96.0%), titanium (IV) isopropoxide [$C_{12}H_{28}O_4Ti$, Sigma Aldrich] (≥97.0%), polyvinylpyrrolidone (PVP) (Alfa Aesar, M.W. 1 300 000), ethanol absolute [$CH_3CH_2OH$, Biosmart] (≥99.9%), acetic acid [$C_2H_4O_2$, VWR Chemicals] (100.0%), 2-methoxyethanol [$C_3H_8O_2$, Oxford] (≥99.0%) and dimethylformamide [$C_3H_7NO$, Sigma Aldrich] (≥99.8%) were used to prepare $Ba_{0.95}Ca_{0.05}Ti_{0.89}Sn_{0.11}O_3$ (BCTSn) and $CoFe_2O_4$ (CFO) electrospinning solutions, and poly(methyl methacrylate) (PMMA) was used as a binder during the poling process for magnetoelectric measurements.

CFO and BCTSn solutions were prepared by sol-gel method as previously reported [20]. Both solutions are mixed together with a molar ration 1:1. The polymer solution was prepared by dissolving PVP in ethanol under vigorous stirring for 4 h. The mixture of CFO and BCTSn sols was added to the PVP solution and stirred continuously to form homogenous CFO and BCTSn polymer solution and then transferred into glass syringe for electrospinning process. The spinning process was carried out at DC voltage of 17 kV, with a 14 cm spacing between the needle tip and the collector and a rate of 0.5 mL $h^{-1}$. The as-spun NFs were dried at 80 °C under vacuum for 12 h before being annealed at 700 °C for 4 h in an air atmosphere.

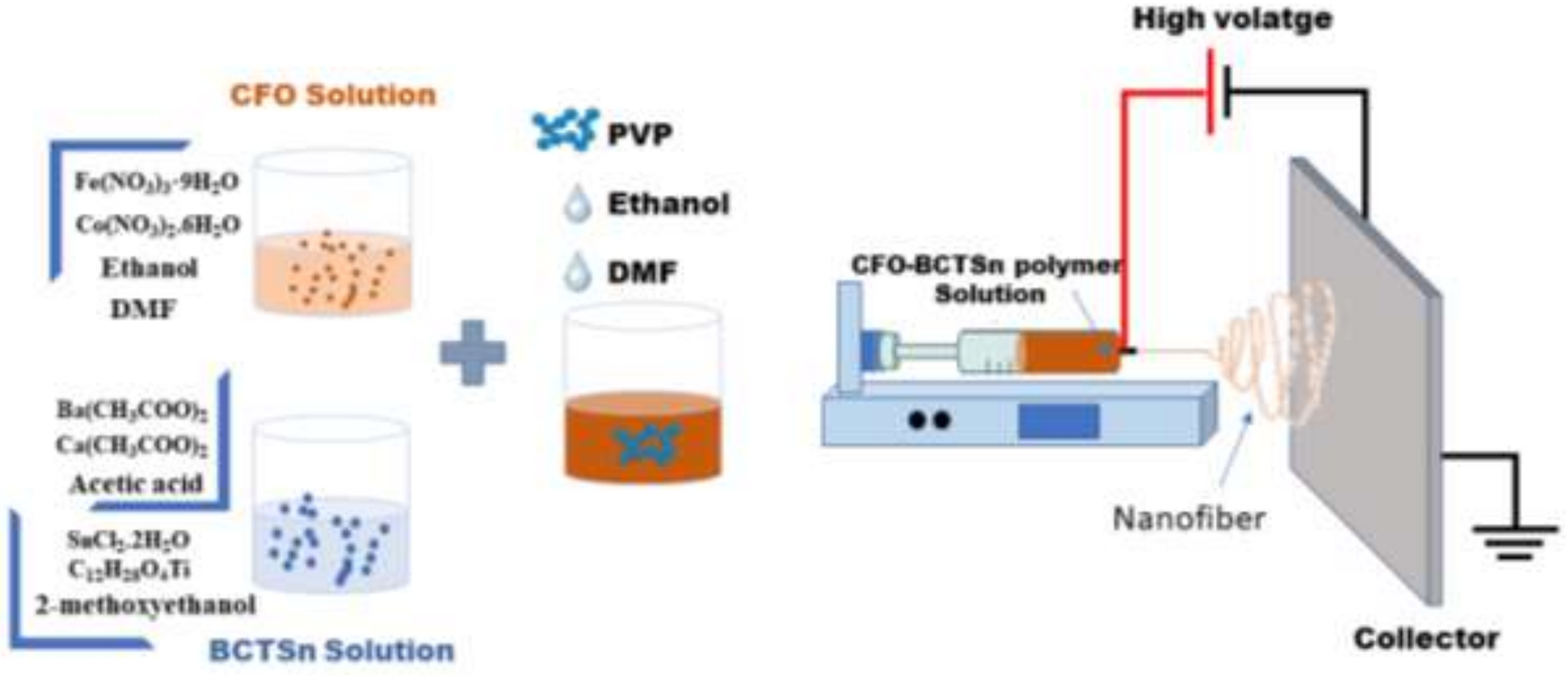

***Figure 1.*** *Schematic diagram of the experimental protocol used to fabricate CFO–BCTSn composite NFs by electrospinning method.*

## 2.2 Characterizations

The morphology and microstructure of CFO–BCTSn composite NFs were observed by a scanning electron microscopy (SEM) in a HELIOS 600 nanolab from Thermofisher and a (JEOL−ARM200F Cold FEG) high-resolution analytical transmission electron microscope (TEM) operating at 200 kV. ImageJ software was used to estimate the fiber diameter distribution by measuring on more than 500 fibers, and the distribution fit was based on the Gaussian function. The crystalline structure of nanofibers was examined at room temperature by x-ray diffraction on a Rigaku (SmartLab SE) with Cu Kα radiation at a scan rate of 2°$min^{-1}$ and an angular scan range (2θ) between 20 and 80°. The macroscopic ferromagnetic property of CFO–BCTSn composite NFs was measured by demonstrating a magnetic hysteresis (M-H) loop at room temperature using a SQUID magnetometer Quantum Design MPMSXL under a magnetic field range of 0-50 kOe for different temperature 2, 150, and 300 K. The local piezoelectric imaging of CFO–BCTSn composite NFs was performed by an atomic force microscope (AFM, Jupiter XR Asylum Research, Oxford Instruments, CA, USA) equipped with a piezo-response force module (PFM). Pt-coated silicon tips with a radius of curvature ~ 10 nm (OMCL-AC240TM-R3, Olympus, Japan) were used for the PFM analysis. The NFs were fixed to the Si substrate by heating at 600 °C for 30 min). More experimental details are given in our recent publication [20]. The images were scanned in dual AC resonance-tracking (DART) vertical and lateral modes. An electrical voltage of 4 V and a frequency of 300 kHz and 800 kHz for vertical and lateral modes were applied, respectively. For magnetoelectric measurements, the CFO–BCTSn composite NFs were embedded in a PMMA matrix. The sample was then divided into two pieces; one piece was coated with silver paste and subjected to poling at 10 kV $cm^{-1}$ for one hour. Subsequently, M-H hysteresis loops were recorded at room temperature for both samples.

## 3. RESULTS AND DISCUSSION

### 3.1 CFO–BCTSn composite NFs: morphology and structure

**Fig. 2a** depicts the SEM image of the as-spun (non-sintered) nanofibers, illustrating the formation of nanofibers with a smooth surface and a non-woven mat morphology. The diameters range from 200 to 500 nm. In **Fig. 2b and c**, SEM images of CFO–BCTSn composite NFs calcined at 700 °C are presented. All fibers exhibit a rough and porous structure, a result of the removal of PVP from the as-spun fibers during sintering, crystallization, and the growth of crystalline phases. The fiber diameters have shrunk to an average of 140 nm, attributed to the removal of PVP after the sintering process (**Fig. 2d**).

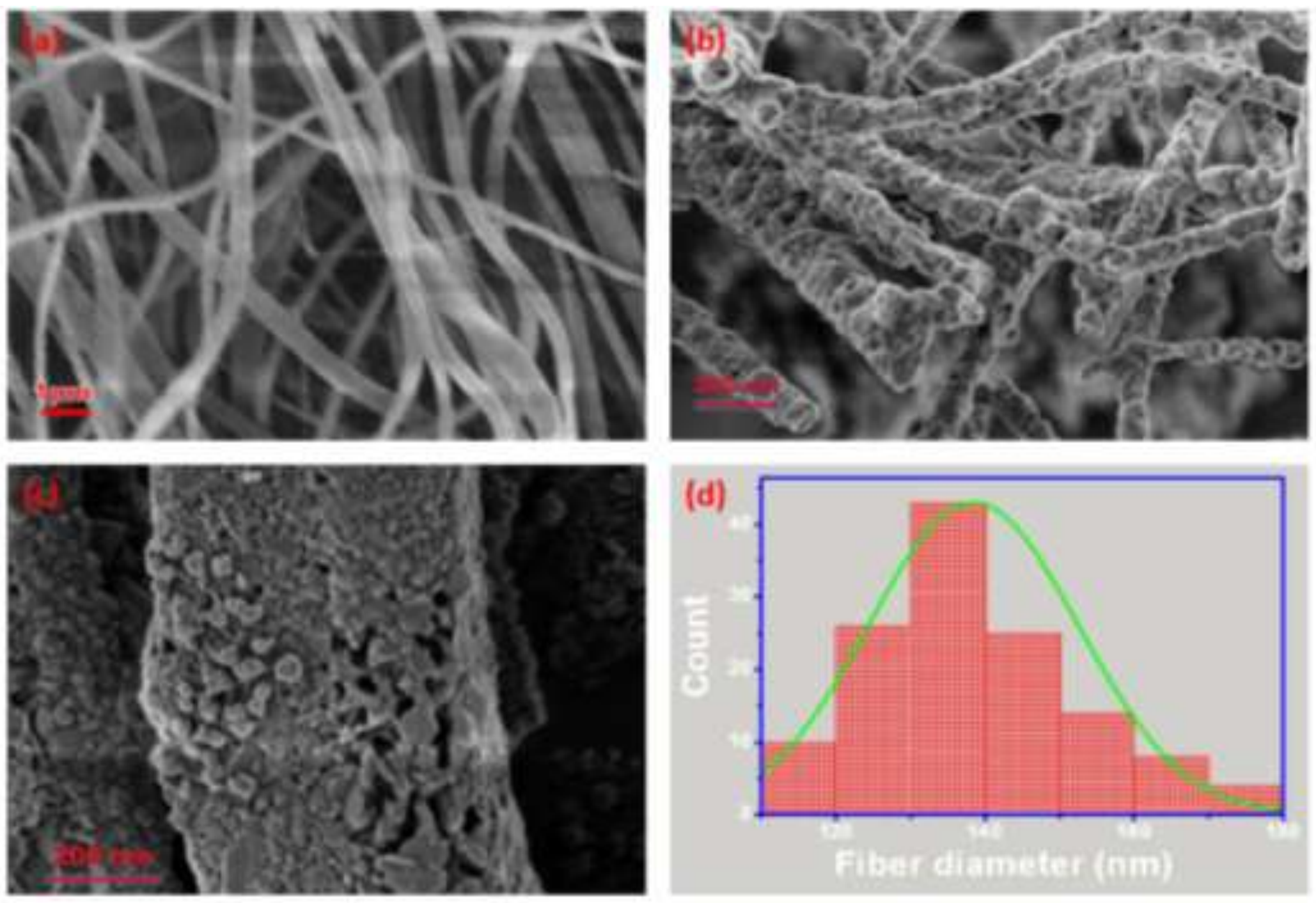

***Figure 2.*** *SEM images: (a) as-spun NFs. (b) calcined CFO–BCTSn composite NFs, (c) surface CFO–BCTSn NF, and (d) fiber diameter distribution.*

The crystalline phases of CFO-BCTSn composite NFs were examined by XRD, as shown in **Fig. 3a**, where two distinct types of diffraction peaks are seen, corresponding to perovskite BCTSn (indicated by +) and spinel CFO phases (indicated by *), respectively, that agree with the standards: Spinel structure (JCPDS No. 01- 1121) and perovskite structure (JCPDS No. 31-0174) [44]. Furthermore, no obvious impurity phases are found in the composite nanofibers. By using the Debye-Scherrer formula to a thorough examination of the peaks broadening of (311) and (100) reflections, the average crystallite size of CFO is estimated to be 9.2 nm, whereas that of BCTSn is roughly 7.5 nm.

**Fig. 3b** shows the Raman spectra of the CFO–BCTSn composite nanofibers recorded in the frequency range of 100 - 750 $cm^{-1}$. The observed Raman modes correspond to the pure BCTSn and CFO phases individually, confirming the coexistence of both ferroic phases without any secondary phases or interactions between them. Specifically, the identified Raman modes at peaks 172, 238, 306, and 517 $cm^{-1}$ are assigned to the BCTSn phase, with the lower frequency modes (<300 $cm^{-1}$) attributed to Ba/Ca–O vibrations, while those at higher wavenumbers correspond to Ti/Sn–O vibrations. For CFO phase, the identified modes are observed at peaks positions 210, 479, 581,621, and 694 $cm^{-1}$, which result from the motion of oxygen ions and both tetrahedral and octahedral sites ions in CFO spinel structure, more detailed explanation is discussed in our recent works [45], [46].

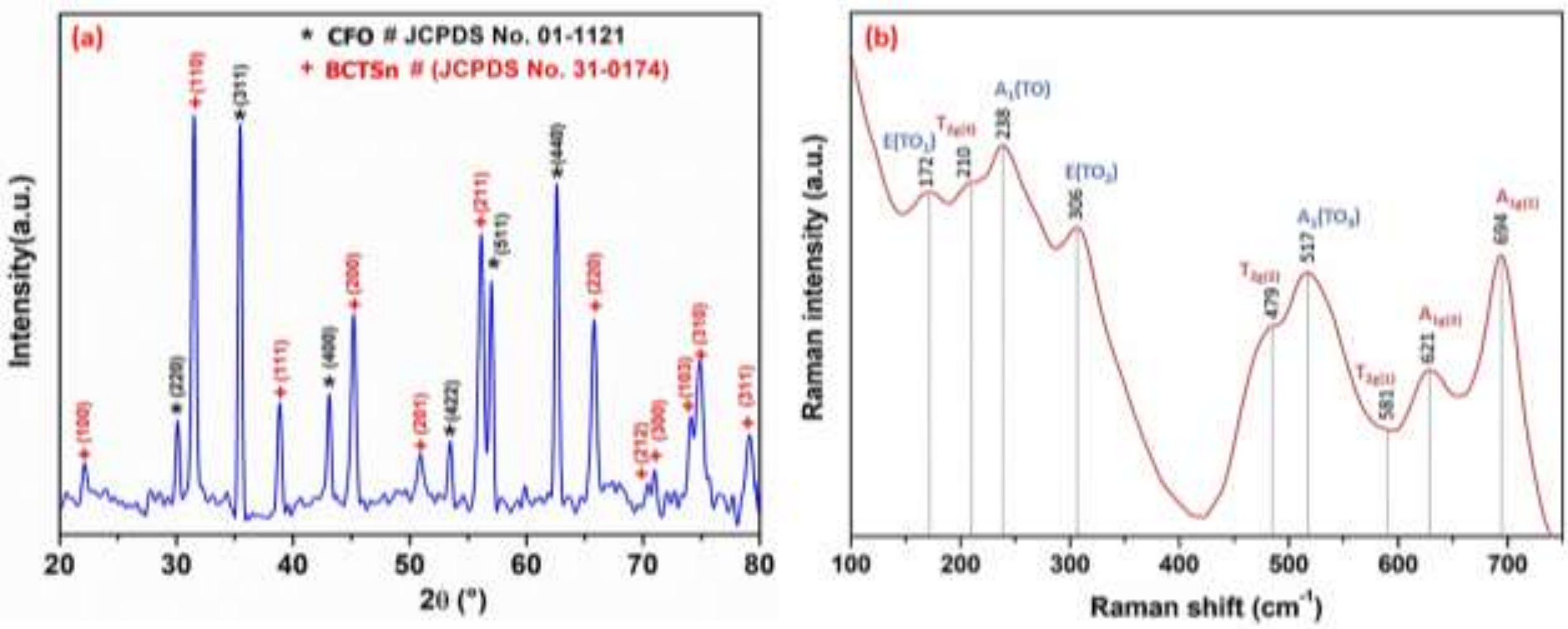

***Figure 3.*** *Room-temperature (a) XRD pattern, (b) Raman spectra of CFO–BCTSn composite NFs.*

**Fig. 4** shows the HRTEM image of the CFO–BCTSn composite NFs. The image reveals lattice fringes with distinct interplanar spacings of 0.406 nm and 0.289 nm, corresponding to the (100) and (110) planes of the BCTSn perovskite phase, respectively. Additionally, spacings of 0.292 nm, 0.478 nm, and 0.261 nm correspond to the (220), (111), and (311) planes of the CFO phase, respectively. It is evident that the CFO and BCTSn phases are coexisted and randomly interconnected. The inset in Fig. 4b shows the selected area electron diffraction (SAED) pattern obtained from the interface of CFO and BCTSn, which displays multiple diffraction rings, suggesting a polycrystalline structure of the CFO–BCTSn composite NFs. This observation confirms the nanoscale nature of the synthesized multiferroic composite.

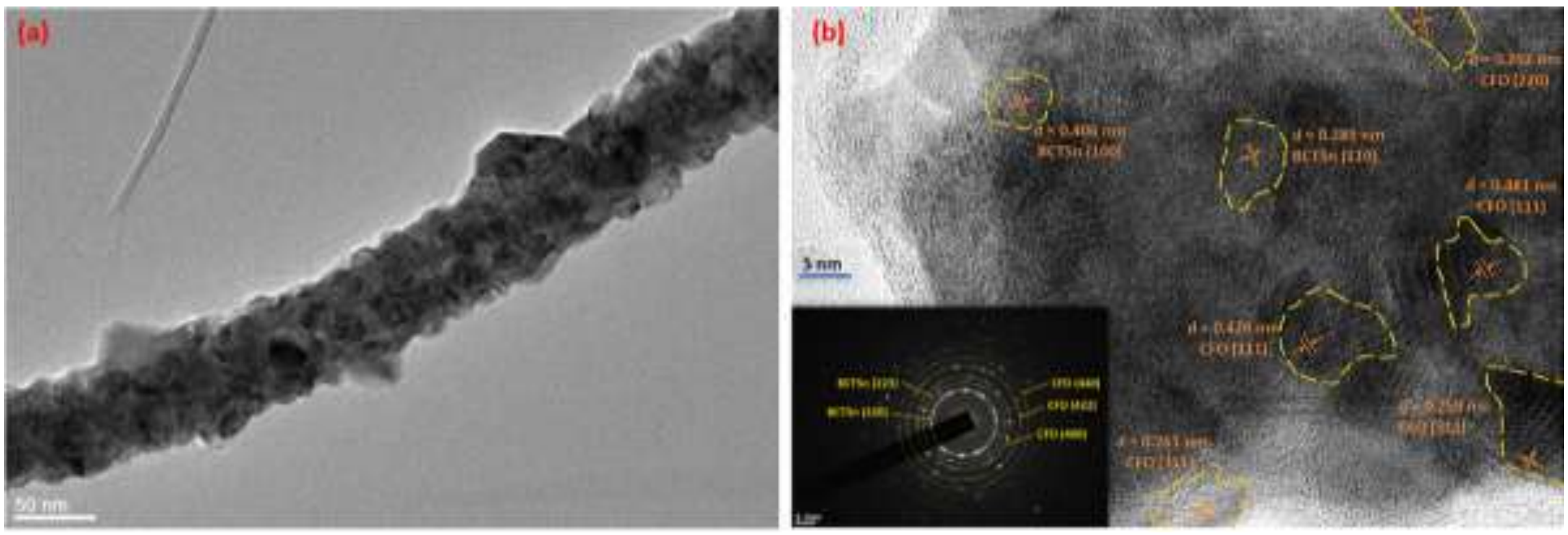

***Figure 4.*** *(a) TEM image; (b) HRTEM image (inset SAED pattern image) of CFO–BCTSn composite NF.*

### 3.2 Magnetic properties of CFO–BCTSn composite NFs

**Fig. 5** displays the M-H loops of CFO–BCTSn composite NFs at temperatures of 2 K, 150 K, and 300 K. These NFs exhibit ferromagnetic behavior attributed to the CFO nanograins, with a saturation magnetization ($M_S$) of approximately 11.5 emu $g^{-1}$ at room temperature. Notably, the M-H curve at low temperature displays a higher $M_S$ (around 15 emu $g^{-1}$ for 150 K and 17 emu $g^{-1}$ for 2 K), compared to room temperature, due to the diminished influence of thermal disturbance on electron spin at lower temperatures. As the total mass of the composite is used to compute the magnetization, the obtained value should be taken as a fraction of the actual magnetization of the magnetic nanograins.

The coercitivity of the composite NFs increases strongly at low temperature, to a value of 12.3 kOe, from a room temperature value of 222 Oe, which can be attributed to the increase of magnetic anisotropy. An important part is the surface/interface anisotropy due to the canted spins at the interface between magnetic and piezoelectric nanoparticles. The irreversibility field, where the magnetizing and demagnetizing branches split, is very large at low temperature (around 35 kOe) and can be identified as the anisotropy field from which an average anisotropy constant K of $1.6x10^6$ emu $cm^{-3}$ is determined (up to 4.5 times higher if a magnetization of 76 emu $g^{-1}$ is used) implying an important contribution of the frozen surface spins. Moreover, the remanence magnetization is only 0.56 of the maximum magnetization at low temperature, sensibly lower than the theoretical value of 0.83 for CFO nanoparticles with cubic anisotropy [47], but very close to the Stoner-Wolfarth uniaxial ansiotropy value of 0.5.

At low temperature, below 150 K, a small jump is visible in the MH loop around zero field, when the field is varied from both positive and negative saturation. This phenomenon, was first observed in CFO nanotubes [48], and was attributed to the low temperature spin reorientation and to domain wall pinning at the interface. The same effect was also observed in CFO nanoparticles and was attributed to the reorientation of the surface spins [49]. The influence of the surface spins on the effective anisotropy is more important at low temperature in our composite NFs (high coercivity and increased $M_r/M_s$ ratio) and diminishes largely at room temperature.

In conclusion, the CFO–BCTSn composite NFs exhibits characteristic hysteresis loops typical of soft magnetic materials at RT, indicating that the magnetism of CFO remains unaffected upon the incorporation of BCTSn in the fibers above 150 K, with an increased contribution to the effective magnetic anisotropy from the interface between the magnetic and piezoelectric nanoparticles at low temperature.

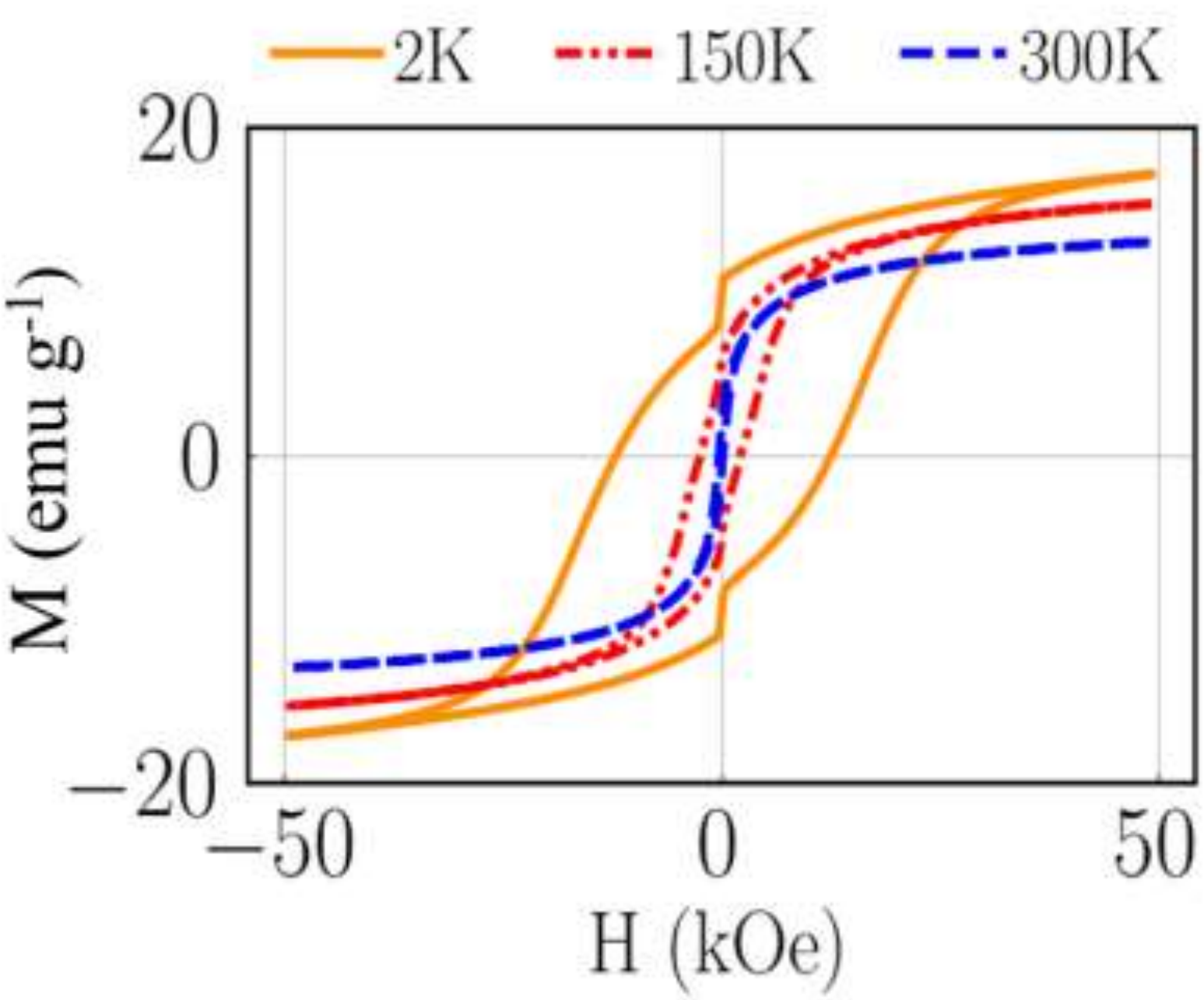

***Figure 5.*** *M-H hysteresis loops at different temperature of CFO–BCTSn composite NFs.*

### 3.3 Piezoelectric properties of CFO–BCTSn composite NFs

**Fig. 6** shows the vertical and lateral PFM experiment results of a single CFO–BCTSn NF on a Si substrate. The AFM height and deflection topographic images in **Fig. 6b** and **c** demonstrate the location and shape of the selected nanofiber on the Si substrate. The piezo-responses obtained in **Fig. 6d** and **f** display different amplitude signals in vertical and lateral directions. A brighter contrast in the lateral PFM amplitude image is clearly visible, compared to a darker contrast in the vertical PFM amplitude image, suggesting a much higher piezoelectric response in the lateral direction. These results suggest that the polarization vector in the NF is probably parallel to the substrate plane, as schematically shown in **Fig. 6a**. An absence of piezoelectric response can be noticed at the bottom of the NF, characterized by the lack of a light spot in lateral and vertical directions (marked by a red arrow in **Fig. 6f**), which may be due to the magnetic phase. The nano-sized areas with the absence of the piezoelectric response are also visible through the whole interior of the investigated NF.

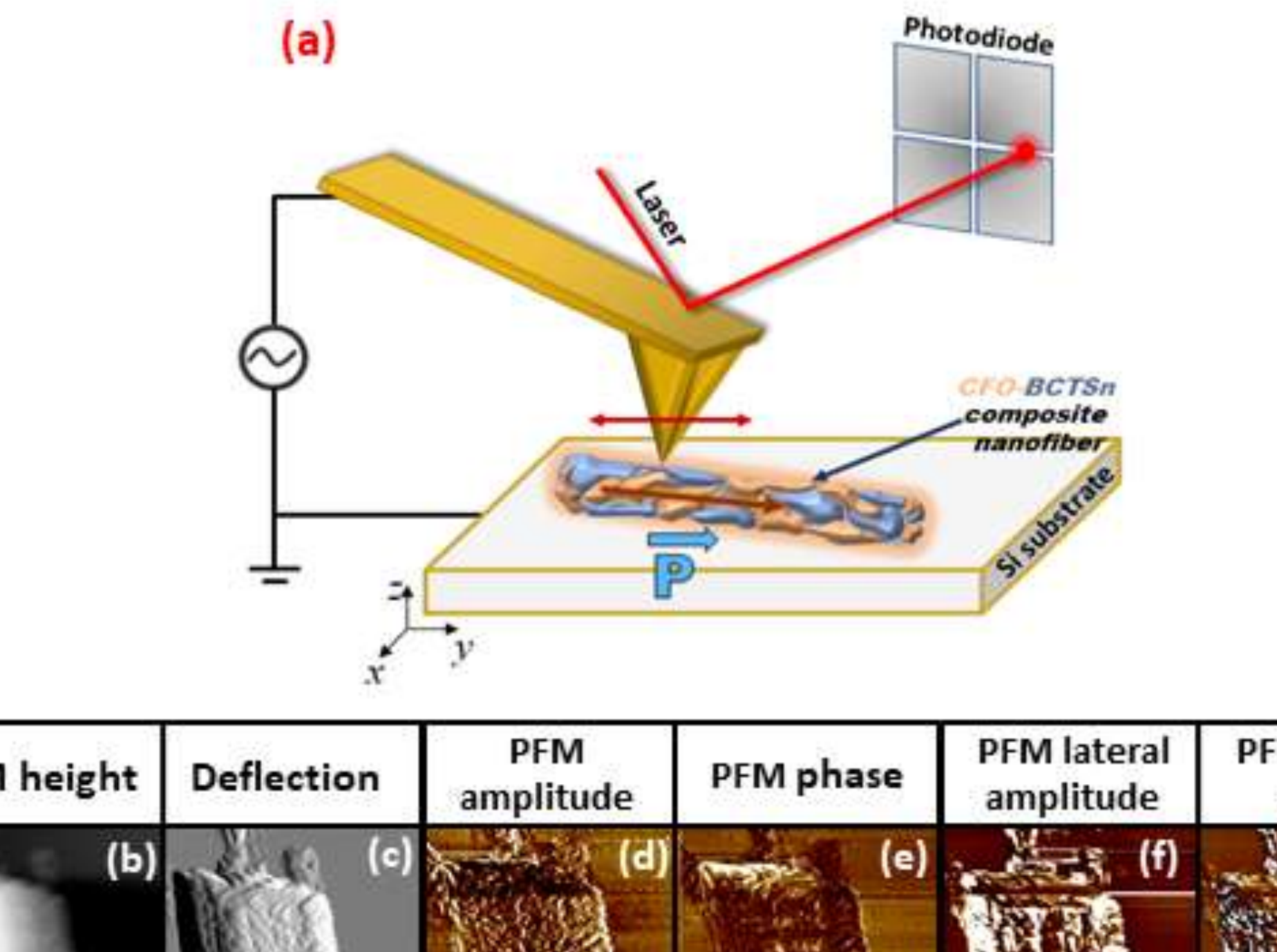

***Figure 6.*** *(a) Schematic illustration of PFM measurement and AFM/PFM images of CFO–BCTSn composite NFs: AFM topography of (b) height and (c) deflection images; vertical PFM (d) amplitude and (e) phase; lateral PFM (f) amplitude and (g) phase images.*

## 3.4 Magnetoelectric properties of CFO–BCTSn composite NFs

The magnetoelectric coupling in the CFO–BCTSn composite fibers was confirmed by analyzing the M-H hysteresis loops of poled and unpoled samples. **Fig. 7** illustrates the M-H hysteresis loops for both samples. The slight increase in saturation magnetization observed in the poled sample suggests the existence of magnetoelectric coupling between the two ferroic phases. Poling the sample aligns the dipole moments along the electric field direction. It is well known that ME coupling involves the elastic interaction between ferroelectric and magnetic domains. The movement of electrical domain walls changes the local strain, which in turn affects the magnetic anisotropy, leading to an increase in magnetization. Other study groups have reported similar results [50], [51], [52]. A quantitative study of the magnetoelectric coupling will be discussed in detail in a forthcoming scientific paper.

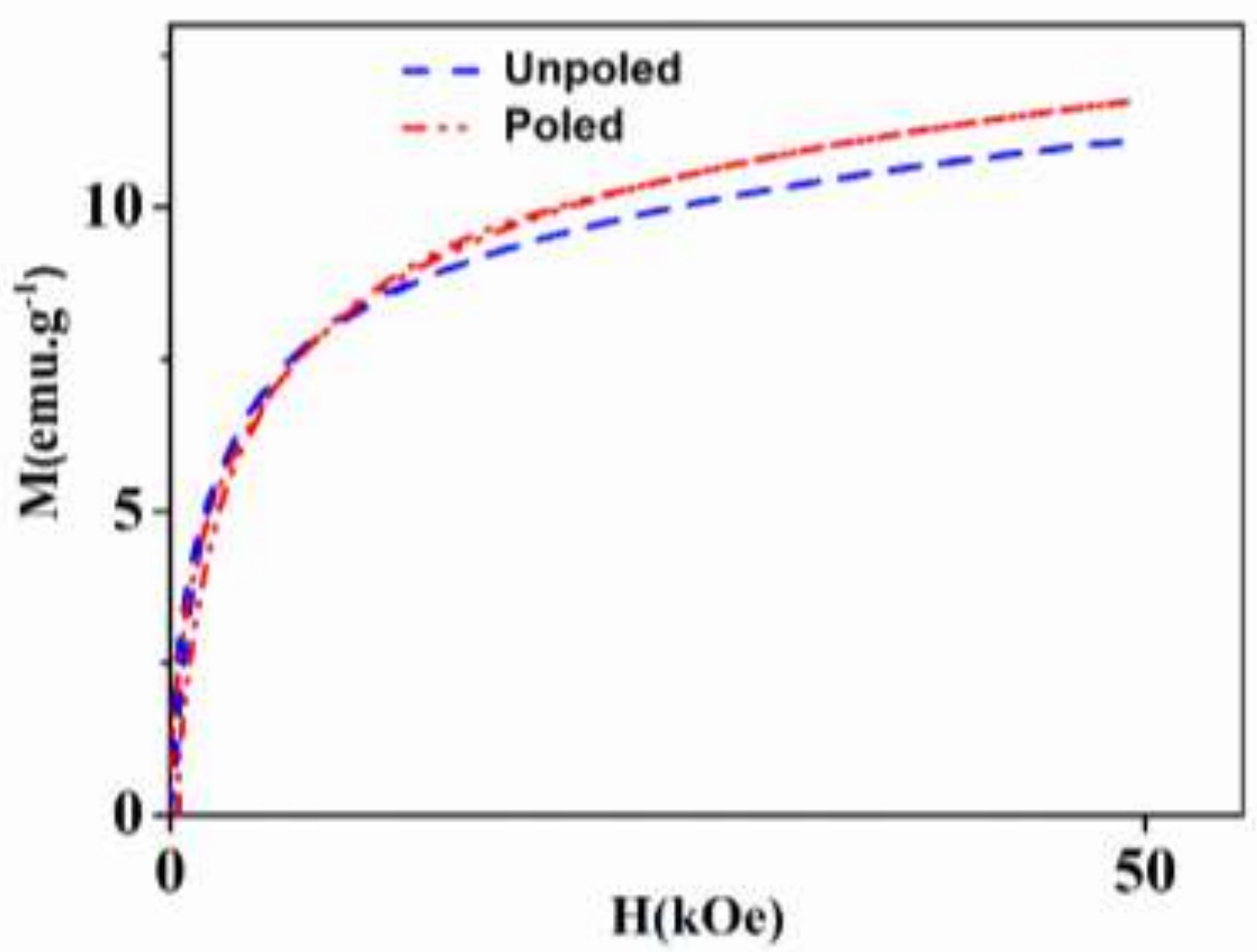

***Figure 7.*** *Magnetic hysteresis loops of both electrically poled and unpoled samples for CFO–BCTSn composite NFs.*

## 4. CONCLUSION

In this paper, multiferroic properties of $CoFe_2O_4$–$Ba_{0.95}Ca_{0.05}Ti_{0.89}Sn_{0.11}O_3$ composite NFs synthesized by sol-gel based electrospinning method were investigated. XRD, Raman spectroscopy, and SAED analysis confirm the presences of both perovskite and spinel structures in the CFO–BCTSn composite NFs. Magnetic properties were studied by M-H hysteresis loops, while piezoelectric properties were verified using PFM. The increase in magnetization values in the M-H loop of poled and unpoled sample proved magnetoelectric coupling between ferrite and ferroelectric phases, suggesting that our CFO-BCTSn NFs composites could be suitable candidates for room-temperature magnetic field sensing applications.

## Acknowledgements

The authors gratefully acknowledge the generous financial support of HORIZON-MSCA-2022-SE H-GREEN (No. 101130520), MSCA-2020-RISE-MELON (No. 872631), and the Slovenian Research Agency (research project N2-0212, research core funding P2-0105) and Jena Cilenšek's technical assistance.